# Deep Learning for Cooperative Radio Signal Classification

Shilian Zheng, Shichuan Chen, and Xiaoniu Yang

## ABSTRACT

Radio signal classification has a very wide range of applications in cognitive radio networks and electromagnetic spectrum monitoring. In this article, we consider scenarios where multiple nodes in the network participate in cooperative classification. We propose cooperative radio signal classification methods based on deep learning for decision fusion, signal fusion and feature fusion, respectively. We analyze the performance of these methods through simulation experiments. We conclude the article with a discussion of research challenges and open problems.

## I. INTRODUCTION

Unknown radio signal classification has a very wide range of applications in cognitive radio networks and electromagnetic spectrum monitoring and management [1][2]. In cognitive radio networks, secondary users can discover the primary users by identifying radio signals, thereby avoiding harmful interference to primary users. In spectrum monitoring, by classifying and identifying each signal in the monitoring frequency band, various illegal user signals can be found, and corresponding measures can be taken to ensure the security of spectrum usage. Traditional radio signal identification methods mainly based on feature engineering to extract signal features. In recent years, deep learning has gained wide attention and application in the classification of radio signals [1]-[5]. The radio signal classification method based on deep learning automatically learns and extracts signal features through deep neural

The authors are with Science and Technology on Communication Information Security Control Laboratory. (Email: lianshizheng@126.com, sicanier@sina.com, yxn2117@126.com)

networks, avoiding a lot of manual analysis and design. In non-cooperative scenarios lacking signal prior information, it shows better performance than traditional signal classification methods.

Most existing deep learning-based radio signal classification methods do not consider multiple receiver scenarios. With the deployment of cognitive radio networks [6] and the emergence of the Internet of Things [7], cooperative classification has become an attractive technical approach to improve classification performance. First, statistically speaking, the signals from multiple nodes with different geographical locations are richer than a single node. Therefore, multi-node cooperation is expected to achieve better classification performance than single node. Secondly, due to the effect of shadowing, the signal received by a certain node may be very weak, which makes it difficult to classify the signal correctly. If it can combine the information of other nodes with better received signal quality, the hidden terminal problem can be avoided. Finally, the improvement of single-node classification performance can be obtained by extending the length of the received signal [8]. However, by cooperative classification, the classification performance can be improved without increasing the signal length, which is beneficial to reduce the computational complexity of a single node.

According to the abstract level of information shared between nodes, cooperative classification can be divided into three categories: decision fusion, signal fusion and feature fusion [9]. In decision fusion, each node sends the local classification result to the fusion node. This method maintains the minimal level of overhead over the network. In signal fusion, each node transmits the observed raw signal data to the fusion node, and the fusion node combines the signals to form a global classifier. This requires minimal computation at the individual nodes with the sacrifice of the largest amount of information exchange overhead over the network. The amount of information overhead of the feature fusion is between the two. In this method, each node transmits the extracted signal features (the dimension is usually much smaller than the raw signal length) to the fusion node which combines these features to globally classify the signal.

In this article, we investigate cooperative radio signal classification methods based on deep learning (convolutional neural networks (CNNs) specifically). We discuss the cooperative radio signal classification problem in the next section. Next we propose three deep learning-based cooperative classification methods for the three fusion scenarios, decision fusion, signal fusion and feature fusion. We use simulation data to evaluate the performance of the proposed methods. Comparison with principle component analysis (PCA)-based feature extraction is also given. We then discuss challenges and research directions for cooperative radio signal classification with deep learning, including tradeoff between computation, overhead and accuracy, countermeasures against adversarial attacks, and open-set cooperative radio signal classification. The final section summarizes the article. To the best of our knowledge, this is the first article that exploits deep learning for cooperative radio signal processing.

## II. PROBLEM DESCRIPTION

Consider cooperative classification with $N$ receiving nodes. Due to the influence of wireless channel propagation and the non-synchronization of the receiving nodes and the transmitter, the signals received by each node have carrier deviation, energy attenuation, noise (additive white Gaussian noise) and other distortions compared with the transmitted signal. Due to the different locations of the receiving nodes, factors such as shadowing cause differences in the quality of the signals received. The purpose of cooperative classification is to integrate information from these different quality signals as much as possible to improve classification accuracy. For simplicity of presentation, the received signal is typically represented by its IQ components (in-phase and quadrature components), i.e., the real and imaginary parts of the received signals. In this article, we consider three cooperative classification scenarios, namely decision fusion, signal fusion and feature fusion.

- Decision fusion. Each node classifies the received signals according to the signal classification algorithm, obtains the local classification result, transmits the classification result to the fusion node. The fusion node combines these decision results to obtain the final global classification.
- Signal fusion. Each node transmits the received raw signal to the fusion node, and the fusion node gives the final classification result based on the fusion algorithm.
- Feature fusion. Each node processes the received signal to obtain the corresponding feature vector, which is transmitted to the fusion node for cooperative classification. The fusion node combines these features and gives the classification result.

In this article, we use deep learning to solve these three cooperative classification problems. Among them, the decision fusion uses deep learning method at the individual nodes, signal fusion uses deep learning method in cooperative classification at the fusion node, and feature fusion uses deep learning method in both feature extraction and fusion. Details of the methods are discussed in the next section.

## III. DEEP LEARNING BASED SOLUTION

### A. BASIC BLOCKS OF CNNS

CNNs have been widely used in recent years to solve radio signal classification problems. Classical CNNs [10] often contain four basic layers: convolutional layer, normalized layer, nonlinear activation layer, and pooling layer. As the number of layers increases, the training error of the traditional CNNs will become larger, that is, the degradation problem will occur. Residual networks [11] are proposed to solve this problem. In this article, we will build CNNs for cooperative radio signal classification based on the basic blocks shown in Fig. 1.

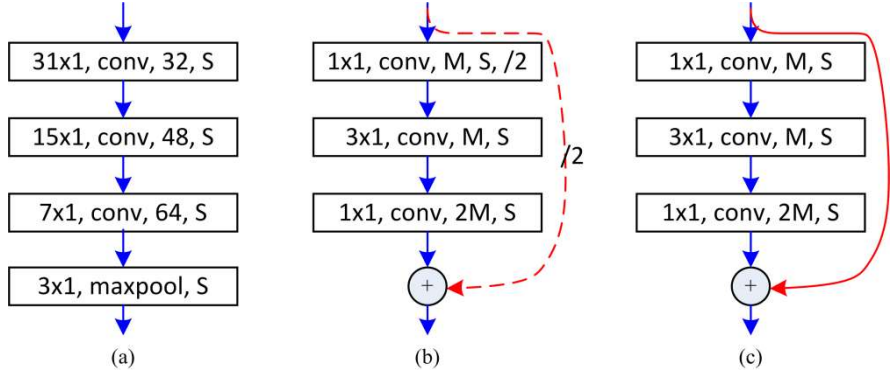

**Figure 1.** Basic network blocks. (a) Con-block，(b) Res-block1(*M*)，(c) Res-block2(*M*)，where *M* denotes the number of convolutional kernels. In the figure, "conv" represents the convolutional layer, the number before which represents the size of the convolution kernel; "S" indicates that the convolution contains padding so that the input and output are of the same size, and "/2" indicates that the downsampling factor is 2, i.e., the output size is reduced to half of the input size; "maxpool" represents maximum pooling. All activations are Rectified Linear Unit (ReLu) functions. A batch normalization layer is included between each convolutional layer and ReLu layer.

## B. DECISION FUSION

The decision fusion radio signal classification based on deep learning is shown in Fig. 2. Each node classifies the received signal based on the same deep neural network and then sends the local decision result to the fusion node. The fusion node combines the received results based on the voting mechanism to give the final classification result.

**Individual signal classification based on CNN.** A CNN (hereinafter referred to as CNN1) for single node signal classification is shown in Fig. 2. The input layer is the IQ representation of the signal (the IQ components are mapped as two channels in the input layer). Unlike the convolution kernel of the CNNs used in the image processing field, the convolution kernel used in this article is a one-dimensional convolution. In order to improve the generalization of the network, we include a Dropout layer after the fully connected layer. Finally, the confidence vector is output through the softmax layer, and the category with the highest probability is used as the single node classification result.

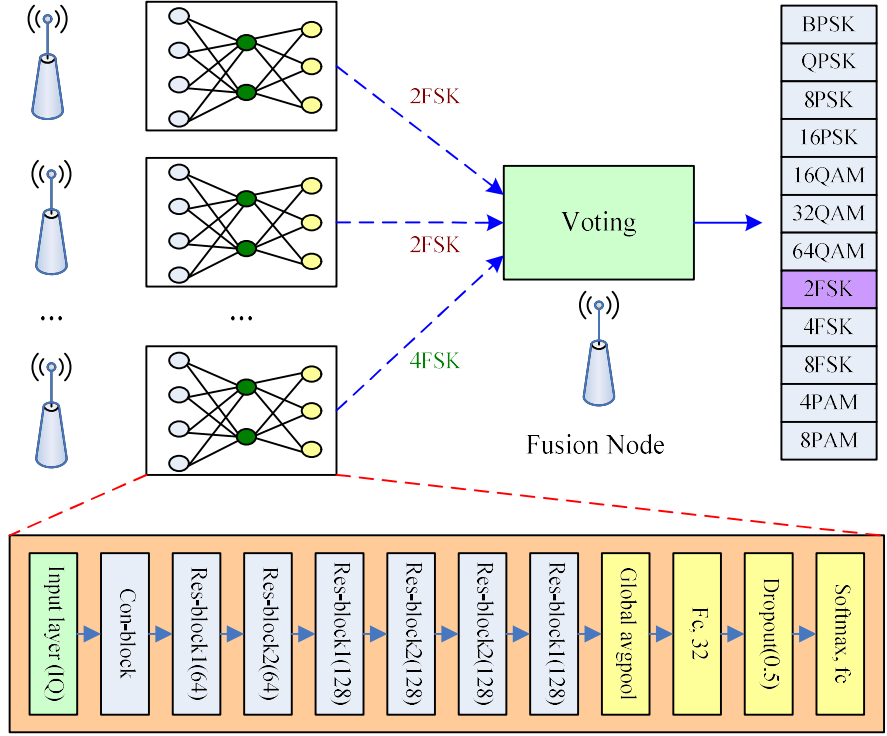

**Figure 2.** Decision fusion cooperative classification. Each node obtains the classification result based on CNN1 and sends the local result to the fusion node. The fusion node gives the final classification result based on the voting mechanism. In the figure, "Global avgpool" represents the global average pooling; "Fc" represents the fully connected layer, the number represents the number of neurons; "Dropout" represents the Dropout layer, the numbers in parentheses indicate the dropout probability; and "Softmax" represents the Softmax layer.

**Fusion based on voting.** The voting mechanism is a widely used fusion method that is based on the principle of majority wins. The fusion node counts the received classification results from each node, and takes the category with the most occurrences as the final classification result.

### C. SIGNAL FUSION

The framework of signal fusion based on deep learning is shown in Fig. 3. Each node receives the radio signals and extracts the signal IQ data and transmits the IQ data to the fusion node. The fusion node combines these data and obtains the classification result based on a CNN (hereafter referred to as CNN2). The CNN2 structure used by the fusion node is basically the same as CNN1, and the main difference lies in the input layer. The input layer of CNN2 combines IQ data from multiple nodes. These data are mapped to different channels of the input layer, so the number of channels in the input layer is 2*N*.

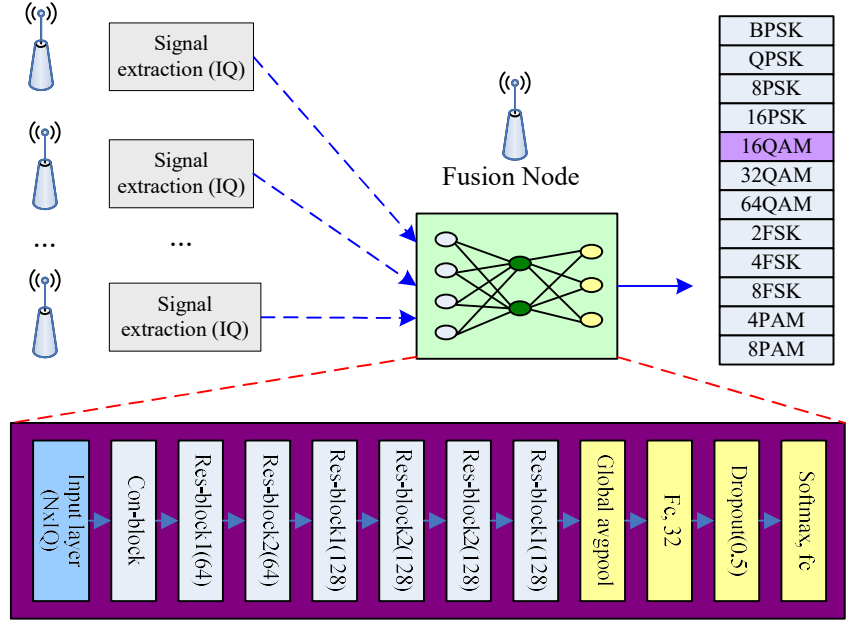

**Figure 3.** Signal fusion cooperative classification. Each node extracts the signal IQ data and transmits it to the fusion node. The fusion node combines these data and obtains the classification result based on CNN2.

### D. FEATURE FUSION

The feature fusion cooperative classification uses a two-stage deep learning mechanism, as shown in Fig. 4. The feature extraction of each node uses the deep learning network CNN1, and the fusion node uses another CNN (hereafter referred to as CNN3).

**Feature extraction based on CNN1.** The network structure adopted by each node to extract features is CNN1. After the CNN1 is trained, we select one of the layers as the signal feature layer. Since the features obtained in deeper layers represent the deeper information of the signal, the feature layer used in this paper is a fully connected layer containing 32 hidden nodes. Therefore, for an IQ signal, the obtained feature is a 32-dimensional vector. The vector will be transmitted to the fusion node for subsequent processing.

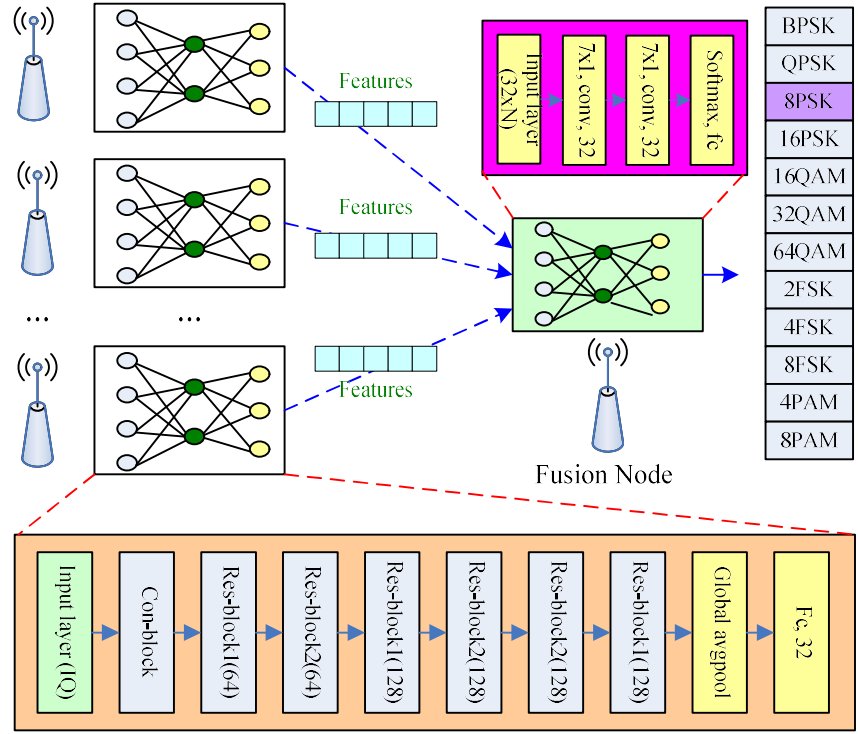

**Figure 4.** Feature fusion cooperative classification. Each node extracts signal features based on CNN1 and transmits the features to the fusion node. The fusion node obtains classification results based on CNN3.

**Feature fusion based on CNN3.** After receiving the feature vectors transmitted by individual nodes, the fusion node performs signal classification based on CNN3 as shown in Fig. 4. The input layer of CNN3 combines feature vectors from these nodes with an input layer dimension of 32x$N$. There are two hidden convolutional layers, all activated by ReLu. Finally, the softmax classification layer outputs the probability of each category, and the index corresponding to the maximum value is the final classification result.

**Training method.** We train CNN1 and CNN3 alone. Firstly, CNN1 is trained by using signal IQ data, and then the features of each signal are extracted using CNN1, and these features are used to train CNN3. It should be noted that the training samples of CNN3 are not from a single node but from multiple nodes in the network.

## IV. SIMULATION RESULTS AND ANALYSIS

### A. DATA GENERATION AND TRAINING SETTINGS

**Data Generation.** In this article, signal data with various modulation types are generated by simulation. The simulation process considers the signal generation process of real communication systems, including pulse shaping, carrier deviation, noise and other factors. The simulation generates signals with 12 modulations: BPSK, QPSK, 8PSK, OQPSK, 2FSK, 4FSK, 8FSK, 16QAM, 32QAM, 64QAM, 4PAM, and 8PAM. Raised cosine pulse shaping filter is used, and the roll-off factor is randomly taken within the range [0.2, 0.7]. The phase deviation is randomly taken, and the normalized carrier frequency offset (relative to the sampling frequency) is randomly taken within the range [-0.1, 0.1]. The signal-to-noise ratio (SNR) ranges from -20 dB to 20 dB with an interval of 2dB. The training and testing samples are as follows.

- For CNN1 training, 1500 samples were generated for each modulation under each SNR, of which two-thirds were used for training and one-third were used for testing. Each signal sample in the generated data contains 64 symbols, and the number of sampling points per symbol is 8, so the signal length is 512.
- For CNN2 training, the same transmitted signal is simulated, and it arrives at the receiving node after different delays, different phase offsets, different frequency deviations, different amplitude attenuations, and different noises. In the simulation, assuming that the time offset between signals received by different nodes is less than one symbol. The SNR difference is at most 20 dB. 1500 samples were generated for each modulation under each SNR, of which two-thirds were used for training and one-third were used for testing.
- In CNN3 training, the signal generation method is similar to the above procedure, except that the generated signals need to be input into the trained

CNN1 network to extract the corresponding 32-dimensional features. The features obtained by different nodes are combined to obtain the training data of CNN3. Similarly, 1500 samples were generated for each modulation under each SNR, of which two-thirds were used for training and one-third were used for testing.

**Training settings.** The training of CNN1, CNN2 and CNN3 adopted the same super-parameter configuration, that is, the initial learning rate is 0.01. The learning rate dropped to one half of the previous learning rate every 10 epochs. The networks were trained for 40 epochs. The training used stochastic gradient descent (SGD) with momentum, with a momentum factor of 0.9. The training was done on the NVIDIA V100 platform.

## B. SIMULATION RESULTS

**Performance of different cooperative methods.** Fig. 5 shows the performance of the three methods at different SNRs, assuming that the signals received by the nodes have the same SNR. It can be seen that the performance of the three cooperative methods is significantly better than that of the non-cooperative approach. In terms of the classification accuracy, the signal fusion performs the best, the decision fusion performs the worst, and the feature fusion lies between the two, as expected. Comparing the results of different number of cooperative nodes, it can be seen that for each doubling of the number of nodes, the feature fusion can obtain an SNR gain of about 1 dB, and the signal fusion can obtain an SNR gain of about 2 dB. For feature fusion, since the original signal length is 512 and the feature dimension is 32, the amount of information exchange overhead is reduced by 16 times. It should be pointed out that for the two-node voting fusion, even if the decision results of the two nodes are inconsistent, the fusion node can only select the decision of one of the nodes as the final classification result. Therefore, the performance of the two-node decision fusion would not be better than the node that performs better.

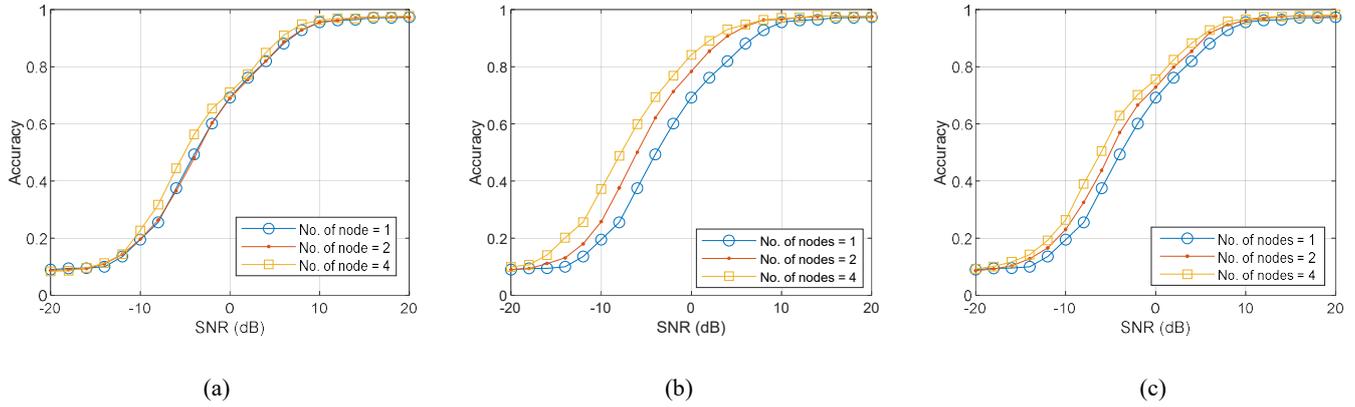

**Figure 5.** Performance of cooperative classification based on deep learning. Signal SNRs of all nodes are assumed to be the same. (a) Decision fusion, (b) signal fusion, and (c) feature fusion.

**Performance with different signal SNRs.** The result given in Fig. 5 is the performance when the signal SNRs received by each node participating in the cooperation is assumed to be the same. However, in most cases, the received signal SNRs will be different due to the different locations of the nodes and the different wireless channel propagation environments. We performed simulation analysis in this case. Fig. 6(a) shows the results of the feature fusion-based cooperative classification experiment. There are 4 nodes participating for cooperative classification, and we assume that the SNR of each node is randomly selected within the range [SNR-ΔSNR, SNR+ΔSNR]. The figure shows the results when the ΔSNR is 0dB, 5dB and 10dB respectively. It can be seen that the larger the ΔSNR is (i.e., the larger the SNR difference), the higher the cooperative classification accuracy is. This is mainly because the greater the range of SNR variation, the higher the probability that a node receives a high SNR signal, and this higher SNR signal is more favorable for signal classification.

**CNN vs. PCA for Feature Fusion.** In the feature fusion method, each node uses the trained CNN1 for signal feature extraction. Another commonly used traditional signal dimension reduction method is the PCA method. Each node can use the PCA to reduce the dimension of the received signal to obtain a 32-dimensional feature vector, and transmit these features to the fusion node for classification. CNN3 model can be trained in a similar manner for cooperative classification. We compare the performance of these two methods. Fig. 6 shows the simulation results. It can be seen that the PCA-based feature fusion cooperative classification performs poorly, even worse than the single-node signal classification method based on CNN1.

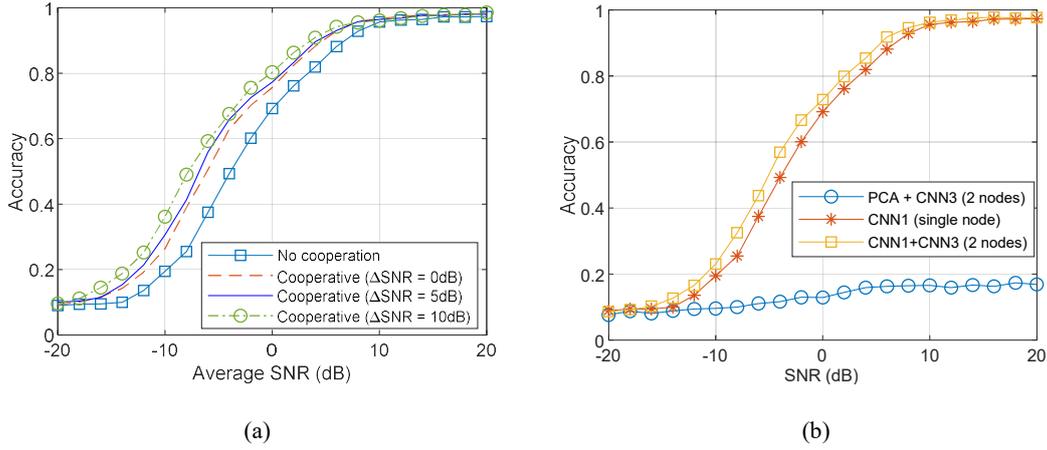

**Figure 6.** Performance of feature fusion-based cooperative classification. (a) Performance with different SNRs and (b) comparison with PCA.

## V. CHALLENGES AND FUTURE DIRECTIONS

### A. TRADEOFF BETWEEN COMPUTATION, OVERHEAD AND ACCURACY

As mentioned earlier, there are three methods for cooperative radio signal: decision fusion, signal fusion, and feature fusion. These methods vary in computational complexity and network communication overhead. Although the decision fusion has the lowest network overhead, the cooperative classification performance gain is not significant enough. Since signal fusion preserves the original raw information, the cooperative classification has the highest accuracy, but the communication burden is also the largest. Feature fusion performs basically between the two. Although PCA, compressed sampling [4], and power spectrum density [3] can be used to reduce the network overhead, they are at the cost of a large loss of classification performance. Therefore, there is still a need to study other signal compression methods. In addition to this, the impact of signal quantization on the performance of cooperative classification also requires further research.

## B. COUNTERMEASURES AGAINST ADVERSARIAL ATTACKS

We know that deep learning models are vulnerable to adversarial attacks [12]. When the fusion node uses deep learning for cooperative signal classification, it also creates possible conditions for malicious attacks. A malicious node in the network can transmit the adversarial samples that is maliciously tampered to the fusion node, thereby affecting the classification result of the fusion node and achieving its attacking purpose. In order to deal with this kind of attacks, it is necessary to study reliable defense methods to ensure the security of cooperative radio signal classification from multiple perspectives such as adversarial sample detection, noise filtering and improving the robustness of the classification neural network model.

## C. OPEN-SET COOPERATIVE RADIO SIGNAL CLASSIFICATION

Most radio signal classification based on deep learning considers closed-set classification scenarios. However, in the actual environment, the system is in an open radio environment, and there are likely to be some untrained radio signal categories. Regardless of single-node radio signal classification or multi-node cooperative radio signal classification, deep learning models trained in closed-set should be required to have the ability to recognize these new categories of signals to avoid false positives which will result in severe degradation of signal classification performance. Therefore, it is necessary to study the cooperative classification in this open-set scenario [13] to improve the robustness of the algorithm in the real-world environment.

## VI. CONCLUSIONS

In this article, we propose cooperative radio signal classification methods based on deep learning to deal with three fusion scenarios: decision fusion, signal fusion and feature fusion. We use the simulation data to evaluate the performance of the proposed methods. Results show that the classification accuracy of the cooperative methods is significantly better than the non-cooperative method. Future work involves tradeoff between computation, overhead and accuracy, countermeasures against adversarial attacks, and open-set cooperative radio signal classification.